\documentclass[preprint,showpacs,preprintnumbers,showkeys,nofootinbib]{revtex4}
\usepackage{amsfonts}
\usepackage{graphicx}
\usepackage{dcolumn}
\usepackage{bm}
\usepackage{tabularx}

\begin{document}

\preprint{}
\title{Geometric Finite Element Discretization of Maxwell Equations in
Primal and Dual Spaces}
\author{Bo He}
\email{he.87@osu.edu}
\author{F. L. Teixeira}
\email{teixeira.5@osu.edu}
\affiliation{ElectroScience Laboratory and Department of Electrical and Computer
Engineering,The Ohio State University, 1320 Kinnear Road, Columbus, OH
43212, USA }
\date{\today }

\begin{abstract}
Based on a geometric discretization scheme for Maxwell equations, we unveil
a mathematical\textit{\ }transformation between the electric field intensity
$E$ and the magnetic field intensity $H$, denoted as Galerkin duality. Using
Galerkin duality and discrete Hodge operators, we construct two system
matrices, $\left[ X_{E}\right] $ (primal formulation) and $\left[ X_{H} %
\right] $ (dual formulation) respectively, that discretize the second-order
vector wave equations. We show that the primal formulation recovers the
conventional (edge-element) finite element method (FEM) and suggests a
geometric foundation for it. On the other hand, the dual formulation
suggests a new (dual) type of FEM. Although both formulations give identical
dynamical physical solutions, the dimensions of the null spaces are
different.
\end{abstract}

\maketitle

\section{Introduction}

The finite element method (FEM), originally developed for structure design
and analysis, is usually based on nodal elements \cite{ZT}. Simply applying
nodal elements to Maxwell equations causes problems such as spurious modes
\cite{Sun}. The use of edge elements is the only reasonable way \cite%
{Bossavit} to remove the spurious modes because the electric field intensity
$E$ is a differential $1$-form with degrees of freedom ($DoFs$) associated
with the edges of a lattice \footnote{%
For high order $1$-forms \cite{Ren}, $DoFs$ of $1$-forms could also
associate with the faces and volumes, but \textit{do not} associate
with the nodes. Pointed out by one reviewer, \ recent work by
Rapetti and Bossavit suggests that $DoFs$ for high order $1$-forms
are still fundamentally associated with segments (small edges).}.

The basic strategy of traditional FEM (Galerkin's method) is to seek the
solution by weighting the residual of the second-order wave equations. Here,
we adopt a different route. Based on a general discretization scheme for
Maxwell equations on irregular lattices, we construct two system matrices in
terms of the electric field intensity $E$ (denoted as primal formulation)
and the magnetic field intensity $H$ (denoted as dual formulation),
respectively. The primal formulation recovers the FEM based on edge
elements, and suggests a geometric foundation for it. On the other hand, the
dual formulation suggests a new (dual) type of FEM. Although both
formulations give identical physical solutions, the dimensions of the null
spaces are different. The connection between the primal formulation and dual
formulation is established via a transformation denoted here as \textit{%
Galerkin duality} (not to be confused with conventional electromagnetic
duality \cite{Balanis}\cite{Chew}).

\section{Discrete Maxwell equations}

Maxwell equations in source-free, three-dimensional (3D) space (in the
Fourier domain) are written in terms of differential forms \cite{Teixeira}%
\cite{Deschamps} as
\begin{equation}
dE=i\omega B,\text{ }dB=0,\text{ }dH=-i\omega D,\text{ }dD=0,
\label{Maxwell}
\end{equation}%
where$\ E$ and $H$ are electric and magnetic field intensity $1$-forms, $D$
and $B$ are electric and magnetic flux $2$-forms, and $d$ is the
(metric-free) exterior derivative operator. We use the convention $%
e^{-i\omega t}$ throughout this paper. Constitutive equations, which include
all metric information, are written in terms of Hodge (star) operators (that
fix an isomorphism between $p$-forms and $\left( 3-p\right) $-forms)

\begin{equation}
D=\star _{\epsilon }E\text{ , }H=\star _{\mu ^{-1}}B.  \label{cons}
\end{equation}

{\Large {\large \ }}By applying basic tools of algebraic topology and a
discrete analog of differential forms, discrete electromagnetic theory can
be constructed from first principles on a general (irregular) primal/dual
lattice (oriented cell-complex)~\cite{Teixeira}. The discrete Maxwell
equations read as \cite{He2}%
\begin{equation}
\left[ d_{curl}\right] \mathbb{E}\mathbb{=}i\omega \mathbb{B}\text{, }\left[
d_{div}\right] \mathbb{B}\mathbb{=}0\text{, }\left[ d_{curl}^{\ast }\right]
\mathbb{H}\mathbb{=}\mathbb{-}i\omega \mathbb{D}\text{, }\left[
d_{div}^{\ast }\right] \mathbb{D}\mathbb{=}0,  \label{incidence}
\end{equation}%
where $\mathbb{E}$, $\mathbb{B}$, $\mathbb{H}$, $\mathbb{D}$ \ are arrays of
$DoFs$ and $\left[ d_{curl}\right] $,$\left[ d_{div}\right] $,$\left[
d_{curl}^{\ast }\right] $, $\left[ d_{div}^{\ast }\right] $ are \textit{%
incidence matrices} that encode the discrete exterior derivatives (discrete
counterparts to the curl and divergence operators, distilled from their
metric structure) on the primal and dual lattice, respectively. Due to the
absence of metric structure, entries of the incidence matrices assume only $%
\left\{ -1,0,1\right\} $ values \cite{Teixeira}.

The discrete Hodge operators can be, in general, written as follows
\begin{equation}
\mathbb{D}=\left[ \star _{\epsilon }\right] \mathbb{E}\text{, }\mathbb{H}=%
\left[ \star _{\mu ^{-1}}\right] \mathbb{B}.  \label{hodgematrix1}
\end{equation}%
One approach to construct the Hodge matrices $\left[ \star _{\epsilon }%
\right] $ and $\left[ \star _{\mu ^{-1}}\right] $ will be discussed in next
Section. The Hodge matrices should be positive definite because Hodge
operators are positive definite (in a Riemannian manifold).

\bigskip

\section{Discrete Hodge operators}

Let $\Omega $ be a $n$-dimensional differentiable manifold and $F^{p}\left(
\Omega \right) $ the space of forms of $p$-degree defined on it. If $\Omega $
is endowed with a metric, then the Hodge operator $\star :\eta \rightarrow
\xi =\star \eta $ \cite{Flanders}\cite{Honan} is defined as a map of $\eta
\in $ $F^{p}\left( \Omega \right) $ to $\xi \in F^{n-p}\left( \Omega \right)
$ such that for any $\psi \in F^{p}\left( \Omega \right) $

\begin{equation}
\int_{\Omega }\psi \wedge \xi =\int_{\Omega }\psi \wedge \star \eta .
\label{hodge}
\end{equation}%
The Hodge operator defines (through a metric) an infinite dimensional inner
product, denoted as $\left( \psi ,\eta \right) $%
\begin{equation}
\left( \psi ,\eta \right) =\int_{\Omega }\psi \wedge \star \eta .
\end{equation}%
For some form $\psi $ we can also define the Hodge square of $\psi $ by%
\begin{equation}
\left( \psi ,\psi \right) =\int_{\Omega }\psi \wedge \star \psi ,
\label{hodgesquare}
\end{equation}%
which is positive when the metric is positive definite. By applying (\ref%
{hodgesquare}) to electric field and magnetic field, one can obtain
constitutive relations in terms of Hodge operators in 3D Euclidean space $%
R^{3}$ as
\begin{eqnarray}
\left( E,E\right) &=&\int_{R^{3}}E\wedge D=\int_{R^{3}}E\wedge \star
_{\epsilon }E,  \label{consED} \\
\left( B,B\right) &=&\int_{R^{3}}B\wedge H=\int_{R^{3}}B\wedge \star _{\mu
^{-1}}B.  \label{consBH}
\end{eqnarray}

Whitney forms \cite{Whitney} are the basic interpolants for discrete
differential forms of various degrees defined over tetrahedra. Whitney forms
can be expressed in term of the barycentric coordinates associated with each
tetrahedron nodes $\left( \zeta _{i},\zeta _{j},\zeta _{k},\zeta _{r}\right)
$ as \cite{Bossavit2}\cite{Bosavit22}
\begin{widetext}
\begin{eqnarray}
w_{i}^{0} &=&\zeta _{i} ,\\
w_{i,j}^{1} &=&\zeta _{i}d\zeta _{j}-\zeta _{j}d\zeta _{i}, \\
w_{i,j,k}^{2} &=&2\left( \zeta _{i}d\zeta _{j}\wedge d\zeta _{k}+\zeta
_{j}d\zeta _{k}\wedge d\zeta _{i}+\zeta _{k}d\zeta _{i}\wedge d\zeta
_{j}\right) ,\\
w_{i,j,k,r}^{3} &=&6\left(
\begin{array}{c}
\zeta _{i}d\zeta _{j}\wedge d\zeta _{k}\wedge d\zeta _{r}-\zeta _{r}d\zeta
_{i}\wedge d\zeta _{j}\wedge d\zeta _{k}\\
+\zeta _{k}d\zeta _{r}\wedge d\zeta _{i}\wedge d\zeta _{j}-\zeta _{j}d\zeta
_{k}\wedge d\zeta _{r}\wedge d\zeta _{i}%
\end{array}%
\right),
\end{eqnarray}%
\end{widetext}(See the appendix for the basis functions over cubes).
Accordingly, we use Whitney $1$-forms as the interpolants for electric field
intensity $1$-form $E$, and Whitney $2$-forms\ as interpolants for the
magnetic flux $2$-form $B$%
\begin{equation}
E=\sum e_{i,j}w_{i,j}^{1},\ B=\sum b_{i,j,k}w_{i,j,k}^{2}.  \label{whitney}
\end{equation}%
Note that the above expansions guarantee tangential continuity of $E$ and
normal continuity of $B$ simultaneously.

Using these basis functions and the Euclidean metric, matrix representations
for the Hodge operators $\star _{\epsilon }$ and $\star _{\mu ^{-1}}$ can be
constructed by combining Eq. (\ref{consED}), Eq. (\ref{consBH}) and Eq. (\ref%
{whitney})
\begin{widetext}
\begin{eqnarray}
\left[ \star _{\epsilon }\right] _{\left\{ \left( i,j\right) ,\left(
\widetilde{i},\widetilde{j}\right) \right\} }
&=&\int_{R^{3}}w_{i,j}^{1}\wedge \star _{\epsilon }w_{\widetilde{i},%
\widetilde{j}}^{1}=\left( w_{i,j}^{1},w_{\widetilde{i},\widetilde{j}%
}^{1}\right)  \nonumber, \\
\left[ \star _{\mu ^{-1}}\right] _{\left\{ \left( i,j,k\right) ,\left(
\widetilde{i},\widetilde{j},\widetilde{k}\right) \right\} }
&=&\int_{R^{3}}w_{i,j,k}^{2}\wedge \star _{\mu ^{-1}}w_{\widetilde{i},%
\widetilde{j},\widetilde{k}}^{2}=\left( w_{i,j,k}^{2},w_{\widetilde{i},%
\widetilde{j},\widetilde{k}}^{2}\right).  \label{hodgematrix}
\end{eqnarray}%
\end{widetext}In the above, matrix entry $\left[ \star _{\epsilon }\right]
_{\left\{ \left( i,j\right) ,\left( \widetilde{i},\widetilde{j}\right)
\right\} }$ comes from edge $\left( i,j\right) $ and edge $\left( \widetilde{%
i},\widetilde{j}\right) $, and matrix entry $\left[ \star _{\mu ^{-1}}\right]
_{\left\{ \left( i,j,k\right) ,\left( \widetilde{i},\widetilde{j},\widetilde{%
k}\right) \right\} }$ comes from face $\left( i,j,k\right) $ and face $%
\left( \widetilde{i},\widetilde{j},\widetilde{k}\right) $. These matrices
denoted as Galerkin's discrete Hodges \cite{Bossavit3}\cite{Bossavit4}, or
simply Galerkin's Hodges.

\section{Primal and dual discrete wave equations}

\subsection{Discrete wave equations\protect\smallskip}

From Eqs.(\ref{incidence}), (\ref{hodgematrix1}) and (\ref{hodgematrix}),
two discrete, second-order vector wave equations can be obtained%
\begin{eqnarray}
\text{ }\left[ d_{curl}^{\ast }\right] \left[ \star _{\mu ^{-1}}\right] %
\left[ d_{curl}\right] \mathbb{E} &\mathbb{=}&\omega ^{2}\left[ \star
_{\epsilon }\right] \mathbb{E},  \label{system1} \\
\left[ d_{curl}\right] \left[ \star _{\epsilon }\right] ^{-1}\left[
d_{curl}^{\ast }\right] \mathbb{H} &\mathbb{=}&\omega ^{2}\left[ \star _{\mu
^{-1}}\right] ^{-1}\mathbb{H},  \label{system2}
\end{eqnarray}%
corresponding to a primal and dual formulation, respectively. These are the
discrete analogs of the curl curl equations%
\begin{eqnarray}
\overrightarrow{\nabla }\frac{1}{\mu }\times \overrightarrow{\nabla }\times
\overrightarrow{E} &=&\omega ^{2}\epsilon \overrightarrow{E},  \label{curl1}
\\
\overrightarrow{\nabla }\frac{1}{\epsilon }\times \overrightarrow{\nabla }%
\times \overrightarrow{H} &=&\omega ^{2}\mu \overrightarrow{H}.
\label{curl2}
\end{eqnarray}%
It can be shown that $\left[ d_{curl}^{\ast }\right] \left[ \star _{\mu
^{-1}}\right] \left[ d_{curl}\right] $ is identical to the conventional
stiffness matrix $\left[ S\right] $ (see Appendix), arising in FEM using
edge elements%
\begin{equation}
\left[ S\right] _{\left\{ \left( i,j\right) ,\left( \widetilde{i},\widetilde{%
j}\right) \right\} }=\int \frac{1}{\mu }\left( \overrightarrow{\nabla }%
\times \overrightarrow{W}_{i,j}^{1}\right) \cdot \left( \overrightarrow{%
\nabla }\times \overrightarrow{W}_{\widetilde{i},\widetilde{j}}^{1}\right)
dV.
\end{equation}%
Moreover, the Hodge matrix $\left[ \star _{\epsilon }\right] $ is identical
to the conventional mass matrix. Hence, the primal formulation recovers the
conventional edge-element FEM and suggests a geometric foundation for it.
For the dual formulation, we can introduce dual stiffness $\left[ S^{\dagger
}\right] $ and mass $\left[ M^{\dagger }\right] $ matrices
\begin{eqnarray}
\left[ S^{\dagger }\right] &=&\left[ d_{curl}\right] \left[ \star _{\epsilon
}\right] ^{-1}\left[ d_{curl}^{\ast }\right] ,  \label{dualstiff} \\
\left[ M^{\dagger }\right] &\mathbb{=}&\left[ \star _{\mu ^{-1}}\right]
^{-1}.  \label{dualmass}
\end{eqnarray}%
This dual formulation has no direct counterpart in traditional FEM. As
discussed next, these two formulations lead to the same dynamic solutions,
but have very different mathematical properties. Note that the Hodge
matrices $\left[ \star _{\epsilon }\right] $ and $\left[ \star _{\mu ^{-1}}%
\right] $ are sparse matrices, but their inverses $\left[ \star _{\epsilon }%
\right] ^{-1}$and $\left[ \star _{\mu ^{-1}}\right] ^{-1}$ are in general
not sparse.

\subsection{Galerkin duality}

\smallskip Galerkin duality is a mathematical transformation between the
above primal and dual formulations. Note that Galerkin duality\textit{\ is
distinct from usual electromagnetic duality~\cite{Balanis}\cite{Chew},} as
illustrated in Table I.

\begin{table*}[tbp]
\caption{Galerkin duality vs. Electromagnetic duality.}
\label{tab:table1}
\par
\begin{tabular}{|l|l|}
\hline
Galerkin duality & Electromagnetic duality \\ \hline
$\overrightarrow{E}\rightarrow \overrightarrow{H}$ $,$ $\overrightarrow{H}%
\rightarrow -\overrightarrow{E}$ & $\overrightarrow{E}\rightarrow
\overrightarrow{H}$ $,$ $\overrightarrow{H}\rightarrow -\overrightarrow{E}$
\\ \hline
PEC$\rightarrow $PEC & PEC$\rightarrow $PMC \\ \hline
Dirichlet BC$\rightarrow $Neumann BC & Dirichlet BC$\rightarrow $Dirichlet BC
\\ \hline
Neumann BC$\rightarrow $Dirichlet BC & Neumann BC$\rightarrow $Neumann BC \\
\hline
\end{tabular}%
\end{table*}
\

Based on Galerkin duality and the discrete Hodge operators introduced
before, we can construct two different system matrices for a given problem%
\begin{eqnarray}
\left[ X_{E}\right] &=&\left[ \star _{\epsilon }\right] ^{-1}\text{ }\left[
d_{curl}^{\ast }\right] \left[ \star _{\mu ^{-1}}\right] \left[ d_{curl}%
\right] , \\
\left[ X_{H}\right] &=&\left[ \star _{\mu ^{-1}}\right] \left[ d_{curl}%
\right] \left[ \star _{\epsilon }\right] ^{-1}\left[ d_{curl}^{\ast }\right]
.
\end{eqnarray}%
Both $\left[ X_{E}\right] $ and $\left[ X_{H}\right] $ encode all discrete
\textit{dynamic} information, and hence produce identical dynamic solutions.
However, their null spaces (associated with zero modes) are very different.
In other words, for a discretization of the same physical system, the
dimensions of the (discrete) zero eigenspaces are different under Galerkin
duality. This can be explained by algebraic properties of discrete Hodge
decomposition, and verified by numerical simulations, as discussed in
Section V.

\subsection{An approach to handle Neumann boundary condition}

Since Dirichlet boundary condition and Neumann boundary condition are
Galerkin dual to each other for some underlying differential equations, we
propose an approach to handle Neumann boundary condition. Consider a
differential equation
\begin{equation}
\Theta \phi =0,  \label{primal}
\end{equation}%
where $\Theta $ is a differential operator and $\phi $ is the unknown
physical quantity, with Neumann boundary condition. By Galerkin duality,
this problem is equivalent to solving
\begin{equation}
\Theta ^{\dagger }\phi ^{\dagger }=0,  \label{dual}
\end{equation}%
with Dirichlet boundary condition. Here $\Theta ^{\dagger }$ is the Galerkin
dual to $\Theta $, and $\phi ^{\dagger }$ is the Galerkin dual to $\phi $.
Note that by using Galerkin duality, we can transform Eq. (\ref{primal})
into Eq. (\ref{dual}), since it may be much easier to handle Dirichlet
boundary conditions than Neumann boundary conditions in some problems. The
function $\phi ^{\dagger }$ can be expanded in terms of basis functions $%
W_{i}^{\dagger }$ (e.g., Whitney forms) as
\begin{equation}
\phi ^{\dagger }=\sum \phi _{i}^{\dagger }W_{i}^{\dagger }.
\end{equation}

\section{\protect\bigskip Examples}

To demonstrate the Galerkin duality, we provide some numerical simulations
for 2D cavity problems in the $\left( x,y\right) $ plane. Both TE and TM
cases are simulated. The finite element meshes for these examples were
generated by using Triangle, a freely available 2D mesh generator \cite%
{Shewchuk}. The angular frequencies of the resonant modes are obtained by
solving the eigenvalue equation (\ref{system1}) (primal formulation) or the
eigenvalue equation (\ref{system2}) (dual formulation). For simplicity, we
set $\epsilon =\mu =1$.

\subsection{Whitney forms in 2D}

The vector proxies of Whitney forms in 2D can be written in term of
barycentric coordinates $\left( \zeta _{i},\zeta _{j},\zeta _{k}\right) $ as%
\begin{widetext}
\begin{eqnarray}
W_{i}^{0} &=&\zeta _{i},\\
\overrightarrow{W}_{i,j}^{1} &=&\zeta _{i}\nabla \zeta _{j}-\zeta _{j}\nabla
\zeta _{i}, \\
W_{i,j,k}^{2} &=&2\left( \zeta _{i}\nabla \zeta _{j}\times \nabla \zeta
_{k}+\zeta _{j}\nabla \zeta _{k}\times \nabla \zeta _{i}+\zeta _{k}\nabla
\zeta _{i}\times \nabla \zeta _{j}\right).
\end{eqnarray}%
\end{widetext}In the above, $W_{i}^{0}$ and $W_{i,j,k}^{2}$ are scalars and $%
\overrightarrow{W}_{i,j}^{1}$ is a vector.

\subsubsection{TE case}

For the TE case, we use $\overrightarrow{W}_{i,j}^{1}$ as the interpolants
for the electric field intensity $\overrightarrow{E}$ and $W_{i,j,k}^{2}$ as
the interpolants for the magnetic flux $B$\smallskip $_{z}$%
\begin{equation}
\overrightarrow{E}=\sum e_{i,j}\overrightarrow{W}_{i,j}^{1},\ B_{z}=\sum
b_{i,j,k}W_{i,j,k}^{2}.
\end{equation}
Galerkin's Hodges become%
\begin{eqnarray}
\left[ \star _{\epsilon }\right] _{\left\{ \left( i,j\right) ,\left(
\widetilde{i},\widetilde{j}\right) \right\} } &=&\int {\epsilon }%
\overrightarrow{W}_{i,j}^{1}\cdot \overrightarrow{W}_{\widetilde{i},%
\widetilde{j}}^{1}dS,  \nonumber \\
\left[ \star _{\mu ^{-1}}\right] _{\left\{ \left( i,j,k\right) ,\left(
\widetilde{i},\widetilde{j},\widetilde{k}\right) \right\} } &=&\int \frac{1}{%
\mu }W_{i,j,k}^{2}\cdot W_{\widetilde{i},\widetilde{j},\widetilde{k}}^{2}dS.
\end{eqnarray}

\subsubsection{TM case}

For the TM case, we use $W_{i}^{0}$ as the interpolants for the electric
field intensity $E_{z}$ and $\overrightarrow{W}_{i,j}^{1}$ as the
interpolants for the magnetic flux \smallskip $\overrightarrow{B}$%
\begin{equation}
E_{z}=\sum e_{i}W_{i}^{0},\ \overrightarrow{B}=\sum b_{i,j}\overrightarrow{W}%
_{i,j}^{1}.
\end{equation}%
Galerkin's Hodges become%
\begin{eqnarray}
\left[ \star _{\epsilon }\right] _{\left\{ i,\widetilde{i}\right\} } &=&\int
{\epsilon }W_{i}^{0}\cdot W_{\widetilde{i}}^{0}dS,  \nonumber \\
\left[ \star _{\mu ^{-1}}\right] _{\left\{ \left( i,j\right) ,\left(
\widetilde{i},\widetilde{j}\right) \right\} } &=&\int \frac{1}{\mu }%
\overrightarrow{W}_{i,j}^{1}\cdot \overrightarrow{W}_{\widetilde{i},%
\widetilde{j}}^{1}dS.
\end{eqnarray}%
The comparison between TE and TM case is illustrated in Table II

\begin{table*}[tbp]
\caption{TE vs.TM.}
\label{tab:table2}%
\begin{tabular}{|l|l|l|}
\hline
& $E$ & $B$ \\ \hline
Degree of differential-form $\left( TE\right) $ & $1$ & $2$ \\ \hline
Degree of differential-form $\left( TM\right) $ & $0$ & $1$ \\ \hline
Element $\left( TE\right) $ & edge & face \\ \hline
Element $\left( TM\right) $ & node & edge \\ \hline
\end{tabular}%
\end{table*}

\subsection{Circular cavity}

Table III and Table IV present the results for TE modes and TM modes of a
circular cavity with radius $a=1.$ The analytical solutions of TE modes are
the zeros of Bessel function derivative $J_{m}^{\prime }\left( x\right) $;
The analytical solutions of TM modes are the zeros of Bessel function $%
J_{m}\left( x\right) $. Note that $TE_{mn}$ and $TM_{mn}$ have a twofold
degeneracy analytically if $m\neq 0$. However, the numerical solutions break
the degeneracy. From the Table III (2D TE modes), we find that the number of
zero modes of primal formulation is equal to the number of internal nodes,
while the number of zero modes of dual formulation is $1$. On the other
hand, from the Table IV (2D TM modes), \ we find that the number of zero
modes of primal formulation is $0$, while the number of zero modes of dual
formulation is $N_{F}-1$. From the last rows of Table III and Table IV, we
conclude that both formulations give the same number of nonzero modes. These
numerical facts, summarized in Table V, will be explained by applying a
discrete Hodge decomposition in next subsection.

\begin{table*}[tbp]
\caption{TE modes (the angular frequencies of the $5$ lowest nonzero modes)
of a circular cavity.}
\label{tab:table3}%
\begin{tabular}{|l|l|l|l|l|}
\hline
Mode $TE_{mn}$ & Primal & Dual & Analytical & Error(\%) \\ \hline
$TE_{11}$ & 1.8493 & 1.8493 & 1.8412 & 0.4416 \\ \hline
$TE_{11}$ & 1.8494 & 1.8494 & 1.8412 & 0.4483 \\ \hline
$TE_{21}$ & 3.0707 & 3.0707 & 3.0542 & 0.5381 \\ \hline
$TE_{21}$ & 3.0708 & 3.0708 & 3.0542 & 0.5412 \\ \hline
$TE_{01}$ & 3.8421 & 3.8421 & 3.8317 & 0.2705 \\ \hline
\# zero modes & 136 & 1 &  &  \\ \hline
\# nonzero modes & 311 & 311 &  &  \\ \hline
\end{tabular}%
\end{table*}

\begin{table*}[tbp]
\caption{TM modes (the angular frequencies of the $5$ lowest nonzero modes)
of a circular cavity.}
\label{tab:table4}%
\begin{tabular}{|l|l|l|l|l|}
\hline
Mode $TM_{mn}$ & Primal & Dual & Analytical & Error(\%) \\ \hline
$TM_{01}$ & 2.4206 & 2.4206 & 2.4048 & 0.6569 \\ \hline
$TM_{11}$ & 3.8883 & 3.8883 & 3.8317 & 1.4758 \\ \hline
$TM_{11}$ & 3.8901 & 3.8901 & 3.8317 & 1.5234 \\ \hline
$TM_{21}$ & 5.2669 & 5.2699 & 5.1356 & 2.5563 \\ \hline
$TM_{21}$ & 5.2694 & 5.2694 & 5.1356 & 2.6050 \\ \hline
\# zero modes & 0 & 311 &  &  \\ \hline
\# nonzero modes & 136 & 136 &  &  \\ \hline
\end{tabular}%
\end{table*}

\begin{table*}[tbp]
\caption{Numerical results of number of modes of TE and\ TM.}
\label{tab:table5}%
\begin{tabular}{|l|l|l|}
\hline
& Primal formulation & Dual formulation \\ \hline
\# zero modes (TE) & $N_{V}^{in}$ & $1$ \\ \hline
\# zero modes (TM) & $0$ & $N_{F}^{{}}-1$ \\ \hline
\# nonzero modes (TE) & $N_{E}^{in}-N_{V}^{in}$ & $N_{F}^{{}}-1$ \\ \hline
\# nonzero modes (TM) & $N_{V}^{in}$ & $N_{E}^{in}-\left( N_{F}-1\right) $
\\ \hline
\end{tabular}%
\end{table*}

\subsection{\protect\bigskip Discrete Hodge decomposition}

In a contractible domain $\Omega $, the Hodge decomposition for a
$p$-form $F^{p}\left( \Omega \right) $ can be written as
\cite{westenholz}
\begin{equation}
F^{p}\left( \Omega \right) =dF^{p-1}\left( \Omega \right) \oplus \delta
F^{p+1}\left( \Omega \right) ,  \label{generalHelholtz}
\end{equation}%
where $\delta $ is the codifferential operator (Hilbert adjoint of $d$). An
arbitrary contractible 2D domain $\Omega $ can be discretized by a general
grid made up of a network of polygons. We will briefly discuss next the
connection between the discrete Hodge decomposition above and the Euler's
formula for a network of polygons (for a more details, see reference \cite%
{He2}).

\subsubsection{2D TE case}

For 2D TE case, applying (\ref{generalHelholtz}) to the electric field
intensity $E$ ($1$-form), we obtain

\begin{equation}
E^{1}=d\phi ^{0}+\delta A^{2},  \label{Helmholtz}
\end{equation}%
where$\ \phi ^{0}$ is a $0$-form and $A^{2}$ is a $2$-form. In Eq. (\ref%
{Helmholtz}) $d\phi ^{0}$ represents the static field and $\delta A^{2}$
represents the dynamic field. We can trace the following correspondence
between Euler's formula for a network of polygons and the Hodge
decomposition~\cite{He2}%
\begin{equation}
\begin{array}[t]{ccc}
N_{E}^{in}-N_{V}^{in} & = & N_{F}-1, \\
\updownarrow \text{ \ \ }\updownarrow &  & \updownarrow \\
E^{1}-d\phi ^{0} & = & \delta A^{2},%
\end{array}
\label{2dte}
\end{equation}%
where $N_{V}^{in}$ is the number of internal vertices, $N_{E}^{in}$ the
number of internal edges and $N_{F}$ the number of faces of a mesh. \

\subsubsection{2D TM case}

For 2D TM case, applying (\ref{generalHelholtz}) to the electric field
intensity $E$ ($0$-form), we obtain

\begin{equation}
E^{0}=\delta A^{1},
\end{equation}%
where$\ A^{1}$ is a $1$-form. We can trace the following correspondence
between Euler's formula for a network of polygons and the Hodge decomposition%
\begin{equation}
\begin{array}[t]{ccc}
N_{V}^{in}-0 & = & \left[ N_{E}^{in}-\left( N_{F}-1\right) \right] , \\
\updownarrow &  & \updownarrow \\
E^{0} & = & \delta A^{1},%
\end{array}
\label{2dtm}
\end{equation}

\subsubsection{Zero modes and nonzero modes}

Eq. (\ref{2dte}) or Eq. (\ref{2dtm}) can be summarized as%
\begin{equation}
L_{1}-L_{2}=R_{1}-R_{2}.  \label{LR}
\end{equation}%
For TE case, we identify%
\begin{equation}
L_{1}=N_{E}^{in},L_{2}=N_{V}^{in},R_{1}=N_{F},R_{2}=1,
\end{equation}%
and for TM case, we identify%
\begin{equation}
L_{1}=N_{V}^{in},L_{2}=0,R_{1}=N_{E}^{in},R_{2}=\left( N_{F}-1\right) .
\end{equation}%
The l.h.s. of Eq. (\ref{LR}) corresponds to the range space of $\left[ X_{E}%
\right] $ while the r.h.s. corresponds to the range space of $\left[ X_{H}%
\right] .$ Furthermore, the $L_{2}$ corresponds to the null space of $\left[
X_{E}\right] $ while $R_{2}$ corresponds to the null space of $\left[ X_{H}%
\right] .$ These results are summarized in Table VI.

\begin{table*}[tbp]
\caption{Null spaces and range spaces of $\left[ X_{E}\right] $ and
$\left[ X_{H}\right] $}
\label{tab:table6}%
\begin{tabular}{|l|l|l|}
\hline
& $\left[ X_{E}\right] $ & $\left[ X_{H}\right] $ \\ \hline
Dim(Null space) (TE) & $N_{V}^{in}$ & $1$ \\ \hline
Dim(Null space) (TM) & $0$ & $N_{F}^{{}}-1$ \\ \hline
Dim(Range space) (TE) & $N_{E}^{in}-N_{V}^{in}$ & $N_{F}^{{}}-1$ \\ \hline
Dim(Range space)(TM) & $N_{V}^{in}$ & $N_{E}^{in}-\left( N_{F}-1\right) $ \\
\hline
\end{tabular}%
\end{table*}

Table VI exactly matches Table V from numerical results. The $DoFs$ of
system matrices $\left[ X_{E}\right] $ and $\left[ X_{H}\right] $ equal the
total number of modes of primal formulation and dual formulation,
respectively. Furthermore, the $DoFs$ in the null space of $\left[ X_{E}%
\right] $ and $\left[ X_{H}\right] $ equal the number of zero modes of
primal formulation and dual formulation, respectively. Finally, the $DoFs$
in the range space of $\left[ X_{E}\right] $ and $\left[ X_{H}\right] $
equal the number of nonzero (dynamic) modes of primal formulation and dual
formulation, respectively. Note that in the case of 2D TE modes (the
electric field intensity $E$ is a $1$-form interpolated by edge elements),
it is a well known fact that the dimension of the null space (\# zero modes
) of $\left[ X_{E}\right] $ is equal to the number of internal nodes \cite%
{He2}\cite{Arnold}\cite{peterson}.

From Eq. (\ref{LR}) (Euler's formula for a network of polygons) it can be
concluded that the dimension of range space of $\left[ X_{E}\right] $ equals
the dimension of range space of $\left[ X_{H}\right] $, as a fundamental
property of discrete Maxwell equations \cite{He2}.

\subsection{Polygonal cavity}

A 2D cavity of arbitrary shape can be approximated by a polygon as the
boundary \cite{He2}. Table VII and Table VIII present the results for TE
modes and TM modes of a polygonal cavity (Fig. 2). These results corroborate
the conclusions summarized by Table V and Table VI. \ Moreover, both systems
matrices $\left[ X_{E}\right] $ and $\left[ X_{H}\right] $ are finite
approximation of the corresponding infinite system. If we use same mesh and
same basis functions, that is, same basic matrices $\left[ d_{curl}\right] $%
, $\left[ d_{curl}^{\ast }\right] $, $\left[ \star _{\mu ^{-1}}\right] $ and
$\left[ \star _{\epsilon }\right] $, the dynamic physical structure encoded
by system matrices $\left[ X_{E}\right] $ and $\left[ X_{H}\right] $ will be%
\textit{\ }identical. Furthermore, if we use same linear solver, the
solutions of both formulations will give \textit{identical} nonzero modes
(dynamic solutions) up to round off errors (see Table VII and VIII).

\begin{table*}[tbp]
\caption{TE modes (the angular frequencies of the $5$ lowest nonzero modes)
of a polygonal cavity.}
\label{tab:table7}%
\begin{tabular}{|l|l|l|}
\hline
Mode No. ($TE$) & Primal formulations & Dual formulation \\ \hline
1 & 2.57359064243139 & 2.57359064243165 \\ \hline
2 & 3.28134124800976 & 3.28134124800987 \\ \hline
3 & 4.32578591632893 & 4.32578591632896 \\ \hline
4 & 5.17188723866480 & 5.17188723866481 \\ \hline
5 & 5.94586993156365 & 5.94586993156362 \\ \hline
\# zero modes & 73 & 1 \\ \hline
\# nonzero modes & 175 & 175 \\ \hline
\end{tabular}%
\end{table*}

\begin{table*}[tbp]
\caption{TM modes (the angular frequencies of the $5$ lowest nonzero modes)
of a polygonal cavity.}
\label{tab:table8}%
\begin{tabular}{|l|l|l|}
\hline
Mode No.($TM$) & Primal formulations & Dual formulation \\ \hline
1 & 4.06172573841605 & 4.06172573841600 \\ \hline
2 & 6.20284873300873 & 6.20284873300876 \\ \hline
3 & 6.85765079948016 & 6.85765079948015 \\ \hline
4 & 8.31632816148913 & 8.31632816148915 \\ \hline
5 & 9.05550834626485 & 9.05550834626483 \\ \hline
\# zero modes & 0 & 175 \\ \hline
\# nonzero modes & 73 & 73 \\ \hline
\end{tabular}%
\end{table*}

\section{ Concluding remarks\ }

Based on Galerkin duality and discrete Hodge operators, we construct two
system matrices, $\left[ X_{E}\right] $ (primal formulation) and $\left[
X_{H}\right] $ (dual formulation) that discretize the wave equations. It can
be shown that the primal formulation recovers conventional (edge-element)
FEM and suggests a geometric foundation for it. On the other hand, the dual
formulation suggests a new (dual) type of FEM. Although both formulations
give identical physical solutions, the null spaces are different. The Hodge
decomposition of the $DoFs$ can be associated with Euler's formula for a
network of polygons in 2D or polyhedra in 3D.

\appendix

\section{Stiffness matrix: \ geometric viewpoint}

Using 3D tetrahedral and cubic elements, respectively, and assuming that the
permeability $\mu $ is constant within each element, we will show that
stiffness matrix $\left[ S\right] $ equals the multiplication of incidences
and Hodge matrices
\begin{equation}
\left[ S\right] =\left[ d_{curl}^{\ast }\right] \left[ \star _{\mu ^{-1}}%
\right] \left[ d_{curl}\right] .
\end{equation}

\subsection{Tetrahedral elements}

From the $DoFs$ for the tetrahedral element (Fig. 3)
\begin{equation}
\mathbb{B}=\left[
\begin{array}[t]{cccc}
b_{1,2,3}^{{}} & b_{1,3,4}^{{}} & b_{1,4,2}^{{}} & b_{2,4,3}^{{}}%
\end{array}%
\right] ^{t},
\end{equation}%
\begin{equation}
\mathbb{E}=\left[
\begin{array}[t]{cccccc}
e_{1,2}^{{}} & e_{1,3}^{{}} & e_{1,4}^{{}} & e_{2,3}^{{}} & e_{4,2}^{{}} &
e_{3,4}^{{}}%
\end{array}%
\right] ^{t},
\end{equation}%
we can construct the incidence matrices $\left[ d_{curl}\right] $ and $\left[
d_{curl}^{\ast }\right] $
\begin{eqnarray}
\left[ d_{curl}\right]  &=&\left[
\begin{array}{cccccc}
1 & -1 & 0 & 1 & 0 & 0 \\
0 & 1 & -1 & 0 & 0 & 1 \\
-1 & 0 & 1 & 0 & 1 & 0 \\
0 & 0 & 0 & -1 & -1 & -1%
\end{array}%
\right] ,\text{ } \\
\left[ d_{curl}^{\ast }\right]  &=&\left[ d_{curl}\right] ^{t}\text{.}
\end{eqnarray}%
In the above, the superscript $t$ stands for transposition. Using the vector
calculus proxies of 3D Whitney 2-form, the Hodge matrix $\left[ \star _{\mu
^{-1}}\right] $ can be calculated as%
\begin{equation}
\left[ \star _{\mu ^{-1}}\right] _{\left\{ \left( i,j,k\right) ,\left(
\widetilde{i},\widetilde{j},\widetilde{k}\right) \right\} }=\int \frac{1}{%
\mu }\overrightarrow{W}_{i,j,k}^{2}\cdot \overrightarrow{W}_{\widetilde{i},%
\widetilde{j},\widetilde{k}}^{2}dV.  \label{hodgebh}
\end{equation}%
Let
\begin{equation}
\left[ G\right] =\left[ d_{curl}^{\ast }\right] \left[ \star _{\mu ^{-1}}%
\right] \left[ d_{curl}\right] ,  \label{stiffness1}
\end{equation}%
which is a $6\times 6$ matrix. The entry of stiffness matrix $\left[ S\right]
$ can be computed as
\begin{widetext}
\begin{eqnarray}
\left[ S\right] _{\left\{ \left( i,j\right) ,\left( \widetilde{i},\widetilde{%
j}\right) \right\} } &=&\int \frac{1}{\mu }\left( \overrightarrow{\nabla }%
\times \overrightarrow{W}_{i,j}^{1}\right) \cdot \left( \overrightarrow{%
\nabla }\times \overrightarrow{W}_{\widetilde{i},\widetilde{j}}^{1}\right) dV
\nonumber \\
&=&\frac{1}{\mu }\left( 2\overrightarrow{\nabla }\zeta _{i}\times
\overrightarrow{\nabla }\zeta _{j}\right) \cdot \left( 2\overrightarrow{%
\nabla }\zeta _{\widetilde{i}}\times \overrightarrow{\nabla }\zeta _{%
\widetilde{j}}\right).  \label{stiffness2}
\end{eqnarray}%
\end{widetext}By comparing each term of matrix (\ref{stiffness1}) with the
corresponding term of matrix (\ref{stiffness2}), such as $\left[ G\right]
_{12}$
\begin{widetext}
\begin{eqnarray}
\left[ G\right] _{12} &=&-\left[ \star _{\mu ^{-1}}\right] _{11}+\left[
\star _{\mu ^{-1}}\right] _{31}+\left[ \star _{\mu ^{-1}}\right] _{12}-\left[
\star _{\mu ^{-1}}\right] _{32} \\
&=&\frac{1}{\mu }\left( 2\overrightarrow{\nabla }\zeta _{1}\times
\overrightarrow{\nabla }\zeta _{2}\right) \cdot \left( 2\overrightarrow{%
\nabla }\zeta _{\widetilde{1}}\times \overrightarrow{\nabla }\zeta _{%
\widetilde{3}}\right),
\end{eqnarray}%
\end{widetext}and $\left[ S\right] _{12}$

\begin{equation}
\left[ S\right] _{12}=\frac{1}{\mu }\left( 2\overrightarrow{\nabla }\zeta
_{1}\times \overrightarrow{\nabla }\zeta _{2}\right) \cdot \left( 2%
\overrightarrow{\nabla }\zeta _{\widetilde{1}}\times \overrightarrow{\nabla }%
\zeta _{\widetilde{3}}\right) ,
\end{equation}%
we obtain \
\begin{equation}
\left[ S\right] =\left[ d_{curl}^{\ast }\right] \left[ \star _{\mu ^{-1}}%
\right] \left[ d_{curl}\right] .
\end{equation}

\subsection{Cubic elements}

Consider a cubic element given in Fig. 4, whose side length is $L$ and whose
center is at $\left( x_{c},y_{c}\right) $. From the $DoFs$ for the cubic
element
\begin{equation}
\mathbb{B}=\left[
\begin{array}[t]{cccccc}
b_{1,4,3,2}^{{}} & b_{5,6,7,8}^{{}} & b_{2,3,7,6}^{{}} & b_{1,5,8,4}^{{}} &
b_{1,2,6,5}^{{}} & b_{3,4,8,7}^{{}}%
\end{array}%
\right] ^{t},
\end{equation}

\begin{widetext}
\begin{equation}
\mathbb{E}=\left[
\begin{array}[t]{cccccccccccc}
e_{1,2}^{{}} & e_{4,3}^{{}} & e_{5,6}^{{}} & e_{8,7}^{{}} & e_{1,4}^{{}} &
e_{5,8}^{{}} & e_{2,3}^{{}} & e_{6,7}^{{}} & e_{1,5}^{{}} & e_{2,6}^{{}} &
e_{4,8}^{{}} & e_{3,7}^{{}}%
\end{array}%
\right] ^{t},
\end{equation}%
\end{widetext}we can construct the incidence matrix $\left[ d_{curl}\right] $
and $\left[ d_{curl}^{\ast }\right] $ for the cubic element%
\begin{widetext}
\begin{equation}
\left[ d_{curl}\right] =\left[
\begin{array}{cccccccccccc}
-1 & 1 &  &  & 1 &  & -1 &  &  &  &  &  \\
&  & 1 & -1 &  & -1 &  & 1 &  &  &  &  \\
&  &  &  &  &  & 1 & -1 &  & -1 &  & 1 \\
&  &  &  & -1 & 1 &  &  & 1 &  & -1 &  \\
1 &  & -1 &  &  &  &  &  & -1 & 1 &  &  \\
& -1 &  & 1 &  &  &  &  &  &  & 1 & -1%
\end{array}%
\right],
\end{equation}%
\end{widetext}%
\begin{equation}
\left[ d_{curl}^{\ast }\right] =\left[ d_{curl}\right] ^{t}.
\end{equation}

The edge elements $\overrightarrow{N}_{i,j}^{1}$ for a cubic element can be
written as \cite{Jin}
\begin{widetext}
\begin{eqnarray}
\overrightarrow{N}_{1,2}^{1} &=&\frac{1}{L^{3}}\left( y_{c}+\frac{L}{2}%
-y\right) \left( z_{c}+\frac{L}{2}-z\right) \widehat{x}  \nonumber, \\
\overrightarrow{N}_{4,3}^{1} &=&\frac{1}{L^{3}}\left( -y_{c}+\frac{L}{2}%
+y\right) \left( z_{c}+\frac{L}{2}-z\right) \widehat{x}  \nonumber, \\
\overrightarrow{N}_{5,6}^{1} &=&\frac{1}{L^{3}}\left( y_{c}+\frac{L}{2}%
-y\right) \left( -z_{c}+\frac{L}{2}+z\right) \widehat{x}  \nonumber, \\
\overrightarrow{N}_{8,7}^{1} &=&\frac{1}{L^{3}}\left( -y_{c}+\frac{L}{2}%
+y\right) \left( -z_{c}+\frac{L}{2}+z\right) \widehat{x}  \nonumber, \\
\overrightarrow{N}_{1,4}^{1} &=&\frac{1}{L^{3}}\left( z_{c}+\frac{L}{2}%
-z\right) \left( x_{c}+\frac{L}{2}-x\right) \widehat{y}  \nonumber, \\
\overrightarrow{N}_{5,8}^{1} &=&\frac{1}{L^{3}}\left( -z_{c}+\frac{L}{2}%
+z\right) \left( x_{c}+\frac{L}{2}-x\right) \widehat{y}  \nonumber, \\
\overrightarrow{N}_{2,3}^{1} &=&\frac{1}{L^{3}}\left( z_{c}+\frac{L}{2}%
-z\right) \left( -x_{c}+\frac{L}{2}+x\right) \widehat{y}  \nonumber, \\
\overrightarrow{N}_{6,7}^{1} &=&\frac{1}{L^{3}}\left( -z_{c}+\frac{L}{2}%
+z\right) \left( -x_{c}+\frac{L}{2}+x\right) \widehat{y}  \nonumber, \\
\overrightarrow{N}_{1,5}^{1} &=&\frac{1}{L^{3}}\left( x_{c}+\frac{L}{2}%
-x\right) \left( y_{c}+\frac{L}{2}-y\right) \widehat{z}  \nonumber, \\
\overrightarrow{N}_{2,6}^{1} &=&\frac{1}{L^{3}}\left( -x_{c}+\frac{L}{2}%
+x\right) \left( y_{c}+\frac{L}{2}-y\right) \widehat{z}  \nonumber, \\
\overrightarrow{N}_{4,8}^{1} &=&\frac{1}{L^{3}}\left( x_{c}+\frac{L}{2}%
-x\right) \left( -y_{c}+\frac{L}{2}+y\right) \widehat{z}  \nonumber, \\
\overrightarrow{N}_{3,7}^{1} &=&\frac{1}{L^{3}}\left( -x_{c}+\frac{L}{2}%
+x\right) \left( -y_{c}+\frac{L}{2}+y\right) \widehat{z}.
\end{eqnarray}%
\end{widetext}The corresponding face elements $\overrightarrow{N}%
_{i,j,k,l}^{2}$ can be constructed as%
\begin{eqnarray}
\overrightarrow{N}_{1,4,3,2}^{2} &=&-\frac{1}{L^{3}}\left( z_{c}+\frac{L}{2}%
-z\right) \widehat{z},  \nonumber \\
\overrightarrow{N}_{5,6,7,8}^{2} &=&\frac{1}{L^{3}}\left( z-z_{c}+\frac{L}{2}%
\right) \widehat{z},  \nonumber \\
\overrightarrow{N}_{2,3,7,6}^{2} &=&\frac{1}{L^{3}}\left( x-x_{c}+\frac{L}{2}%
\right) \widehat{x},  \nonumber \\
\overrightarrow{N}_{1,5,8,4}^{2} &=&-\frac{1}{L^{3}}\left( x_{c}+\frac{L}{2}%
-x\right) \widehat{x},  \nonumber \\
\overrightarrow{N}_{1,2,6,5}^{2} &=&-\frac{1}{L^{3}}\left( y_{c}+\frac{L}{2}%
-y\right) \widehat{y},  \nonumber \\
\overrightarrow{N}_{3,4,8,7}^{2} &=&\frac{1}{L^{3}}\left( y-y_{c}+\frac{L}{2}%
\right) \widehat{y}.
\end{eqnarray}%
The Hodge matrix $\left[ \star _{\mu ^{-1}}\right] $ can be calculated as%
\begin{equation}
\left[ \star _{\mu ^{-1}}\right] _{\left\{ \left( i,j,k,l\right) ,\left(
\widetilde{i},\widetilde{j},\widetilde{k},\widetilde{l}\right) \right\}
}=\int \frac{1}{\mu }\overrightarrow{W}_{i,j,k,l}^{2}\cdot \overrightarrow{W}%
_{\widetilde{i},\widetilde{j},\widetilde{k},\widetilde{l}}^{2}dV,
\end{equation}%
\begin{equation}
\left[ \star _{\mu ^{-1}}\right] =\frac{1}{6L\mu }\left[
\begin{array}{cccccc}
2 & -1 & 0 & 0 & 0 & 0 \\
-1 & 2 & 0 & 0 & 0 & 0 \\
0 & 0 & 2 & -1 & 0 & 0 \\
0 & 0 & -1 & 2 & 0 & 0 \\
0 & 0 & 0 & 0 & 2 & -1 \\
0 & 0 & 0 & 0 & -1 & 2%
\end{array}%
\right] .
\end{equation}

\bigskip Let $c=\frac{1}{6L\mu }$. The matrix $\left[ G\right] $ can be
computed as

\begin{widetext}
\begin{eqnarray}
\left[ G\right] &=&\left[ d_{curl}^{\ast }\right] \left[ \star _{\mu ^{-1}}%
\right] \left[ d_{curl}\right]  \nonumber \\
&=&c\left[
\begin{array}{cccccccccccc}
4 & -1 & -1 & -2 & -2 & -1 & 2 & 1 & -2 & 2 & -1 & 1 \\
-1 & 4 & -2 & -1 & 2 & 1 & -2 & -1 & -1 & 1 & -2 & 2 \\
-1 & -2 & 4 & -1 & -1 & -2 & 1 & 2 & 2 & -2 & 1 & -1 \\
-2 & -1 & -1 & 4 & 1 & 2 & -1 & -2 & 1 & -1 & 2 & -2 \\
-2 & 2 & -1 & 1 & 4 & -1 & -1 & -2 & -2 & -1 & 2 & 1 \\
-1 & 1 & -2 & 2 & -1 & 4 & -2 & -1 & 2 & 1 & -2 & -1 \\
2 & -2 & 1 & -1 & -1 & -2 & 4 & -1 & -1 & -2 & 1 & 2 \\
1 & -1 & 2 & -2 & -2 & -1 & -1 & 4 & 1 & 2 & -1 & -2 \\
-2 & -1 & 2 & 1 & -2 & 2 & -1 & 1 & 4 & -1 & -1 & -2 \\
2 & 1 & -2 & -1 & -1 & 1 & -2 & 2 & -1 & 4 & -2 & -1 \\
-1 & -2 & 1 & 2 & 2 & -2 & 1 & -1 & -1 & -2 & 4 & -1 \\
1 & 2 & -1 & -2 & 1 & -1 & 2 & -2 & -2 & -1 & -1 & 4%
\end{array}%
\right]  \nonumber. \\
&&  \label{stiffness3}
\end{eqnarray}
\end{widetext}

Using the formula
\begin{equation}
\left[ S\right] _{\left\{ \left( i,j\right) ,\left( \widetilde{i},\widetilde{%
j}\right) \right\} }=\int \frac{1}{\mu }\left( \overrightarrow{\nabla }%
\times \overrightarrow{N}_{i,j}^{1}\right) \cdot \left( \overrightarrow{%
\nabla }\times \overrightarrow{N}_{\widetilde{i},\widetilde{j}}^{1}\right)
dV,
\end{equation}%
the stiffness matrix $\left[ S\right] $ can be computed as%
\begin{widetext}
\begin{eqnarray}
\left[ S\right] &=&c\left[
\begin{array}{cccccccccccc}
4 & -1 & -1 & -2 & -2 & -1 & 2 & 1 & -2 & 2 & -1 & 1 \\
-1 & 4 & -2 & -1 & 2 & 1 & -2 & -1 & -1 & 1 & -2 & 2 \\
-1 & -2 & 4 & -1 & -1 & -2 & 1 & 2 & 2 & -2 & 1 & -1 \\
-2 & -1 & -1 & 4 & 1 & 2 & -1 & -2 & 1 & -1 & 2 & -2 \\
-2 & 2 & -1 & 1 & 4 & -1 & -1 & -2 & -2 & -1 & 2 & 1 \\
-1 & 1 & -2 & 2 & -1 & 4 & -2 & -1 & 2 & 1 & -2 & -1 \\
2 & -2 & 1 & -1 & -1 & -2 & 4 & -1 & -1 & -2 & 1 & 2 \\
1 & -1 & 2 & -2 & -2 & -1 & -1 & 4 & 1 & 2 & -1 & -2 \\
-2 & -1 & 2 & 1 & -2 & 2 & -1 & 1 & 4 & -1 & -1 & -2 \\
2 & 1 & -2 & -1 & -1 & 1 & -2 & 2 & -1 & 4 & -2 & -1 \\
-1 & -2 & 1 & 2 & 2 & -2 & 1 & -1 & -1 & -2 & 4 & -1 \\
1 & 2 & -1 & -2 & 1 & -1 & 2 & -2 & -2 & -1 & -1 & 4%
\end{array}%
\right]  \nonumber. \\
&&  \label{stiffness4}
\end{eqnarray}
\end{widetext}

Comparison of Eq.(\ref{stiffness3}) and Eq.(\ref{stiffness4}) gives the
following identity \
\begin{equation}
\left[ S\right] =\left[ d_{curl}^{\ast }\right] \left[ \star _{\mu ^{-1}}%
\right] \left[ d_{curl}\right] .
\end{equation}%
The above proof can be straightforwardly extended to the rectangular brick
element whose side lengths are $\left( L_{x},L_{y},L_{z}\right) $.\pagebreak

Fig.1. The mesh has 178 vertices (136 internal vertices), 447 internal
edges, and 312 triangles.

\bigskip

Fig.2. The coordinates of the vertices of the polygon are $\left( 0,0\right)
,\left( 1,0\right) ,\left( 1.4,0.4\right) ,\left( 1.3,1.0\right) ,\left(
0.8,1.2\right) ,\left( 0.3,0.9\right) $. The mesh has 105 vertices (73
internal vertices), 248 internal edges, and 176 triangles.

\bigskip

Fig.3. Oriented tetrahedral element.

\bigskip

Fig. 4. Oriented cubic element.

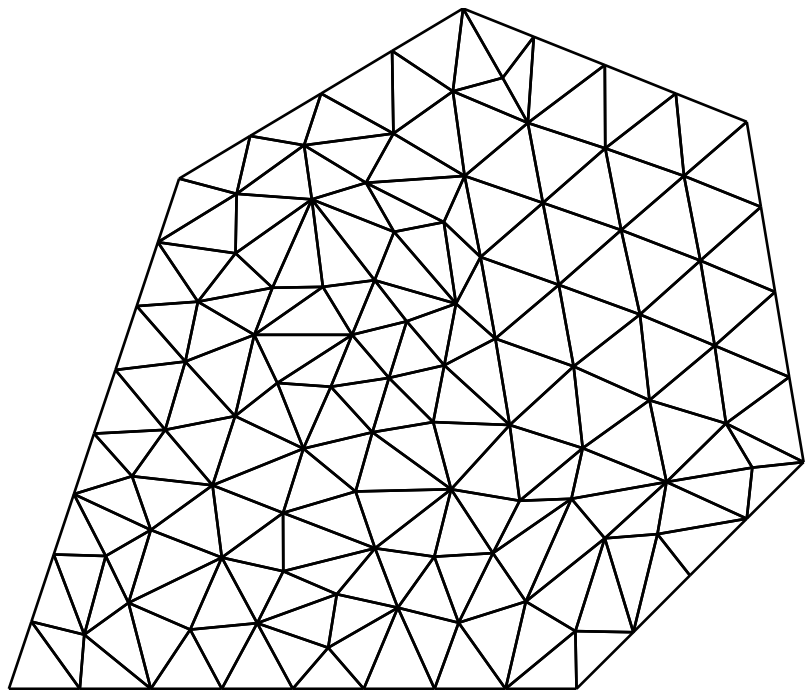
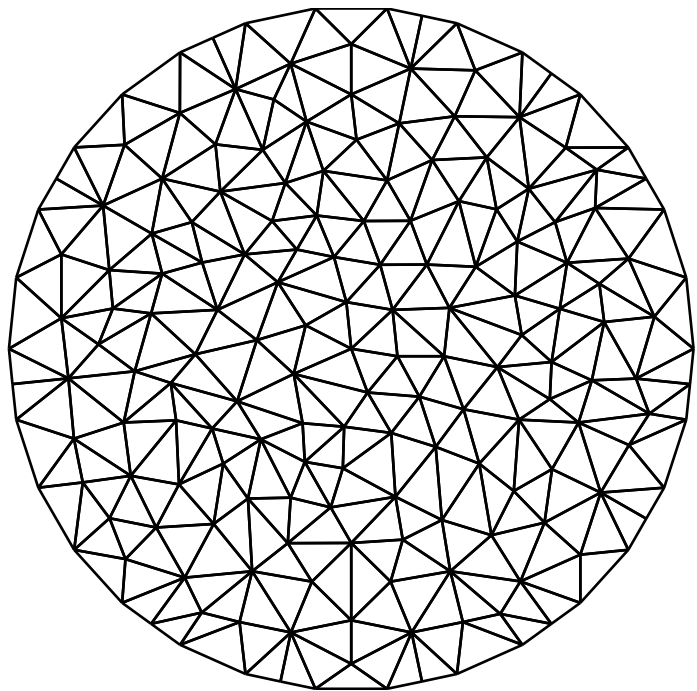
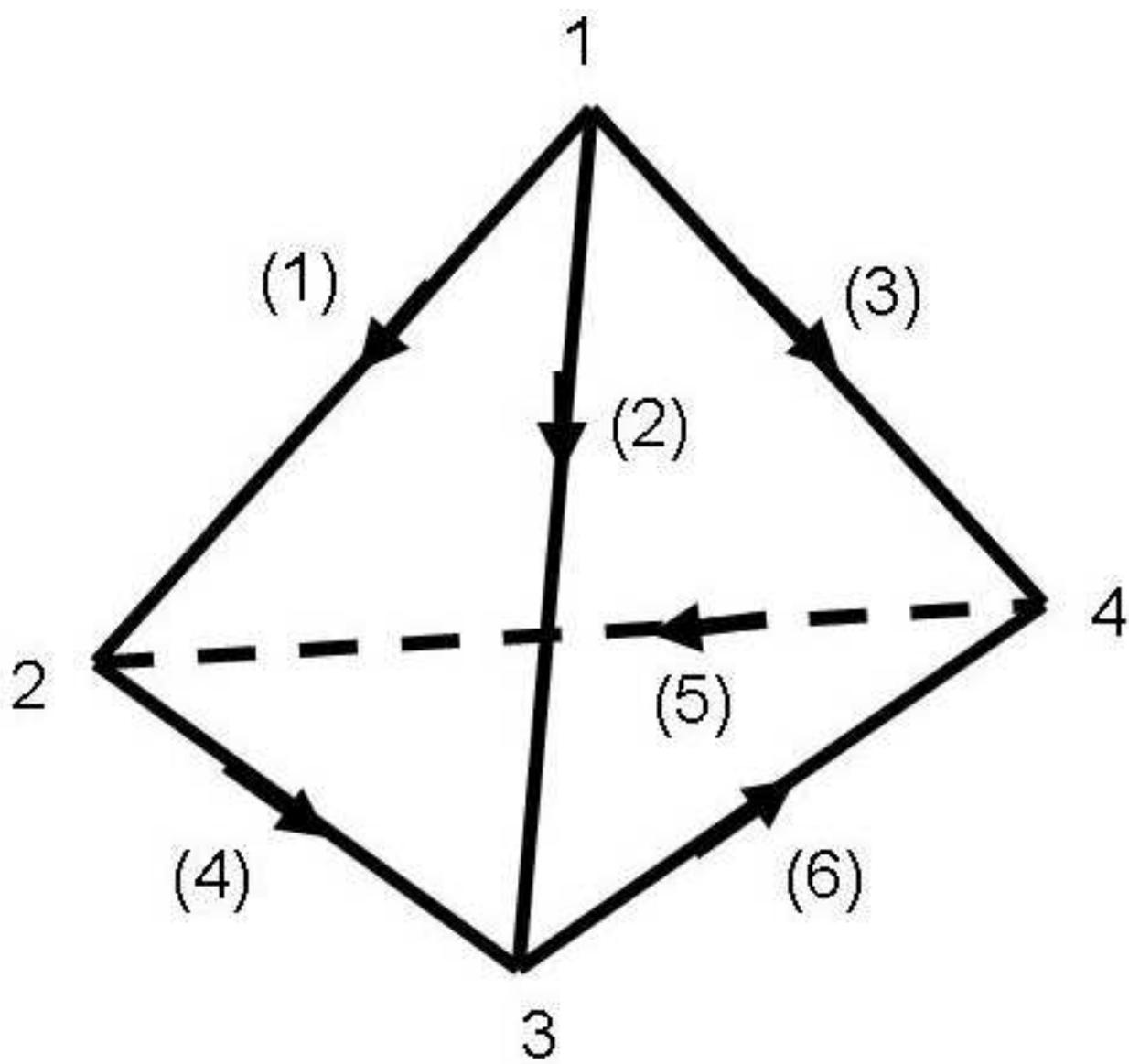
\end{document}